\documentclass[journal=nalefd,manuscript=letter]{achemso}
\usepackage{subfig,graphicx}
\usepackage[T1]{fontenc} 
\usepackage{bbm}
\usepackage{bm}
\usepackage{float}
\usepackage{ulem}
\usepackage{amsmath}
\usepackage{amssymb}
\usepackage{empheq}
\usepackage{graphicx}
\usepackage{mathrsfs}
\usepackage{amsfonts}
\usepackage{amsthm}
\usepackage{color}
\usepackage{bigints}
\usepackage{txfonts}
\usepackage{hyperref}
\usepackage{pifont}
\usepackage{lettrine}

\title[]
{Topological equivalence of stripy states and skyrmion crystals}

\author{X. R. Wang}
\affiliation[The Hong Kong University of Science and Technology]
{Physics Department, The Hong Kong University of Science 
and Technology (HKUST), Clear Water Bay, Kowloon, Hong Kong}
\alsoaffiliation[HKUST Shenzhen Research Institute]
{HKUST Shenzhen Research Institute, Shenzhen 518057, China}
\email{phxwan@ust.hk}
\author{Xu-Chong Hu}
\affiliation[The Hong Kong University of Science and Technology]
{Physics Department, The Hong Kong University of Science 
and Technology (HKUST), Clear Water Bay, Kowloon, Hong Kong}
\alsoaffiliation[HKUST Shenzhen Research Institute]
{HKUST Shenzhen Research Institute, Shenzhen 518057, China}
\author{Zhou-Zhou Sun}
\affiliation[South China Business College]
{South China Business College, Guangdong University of 
Foreign Studies, Guangzhou 510545, China}
\alsoaffiliation[HKUST Shenzhen Research Institute]
{HKUST Shenzhen Research Institute, Shenzhen 518057, China}

\abbreviations{SkX}
\keywords{Nano-magnetism, chiral magnets, topological physics, 
skyrmion crystals, helical states}

\begin{document}

\begin{tocentry}

\includegraphics[width=\columnwidth]{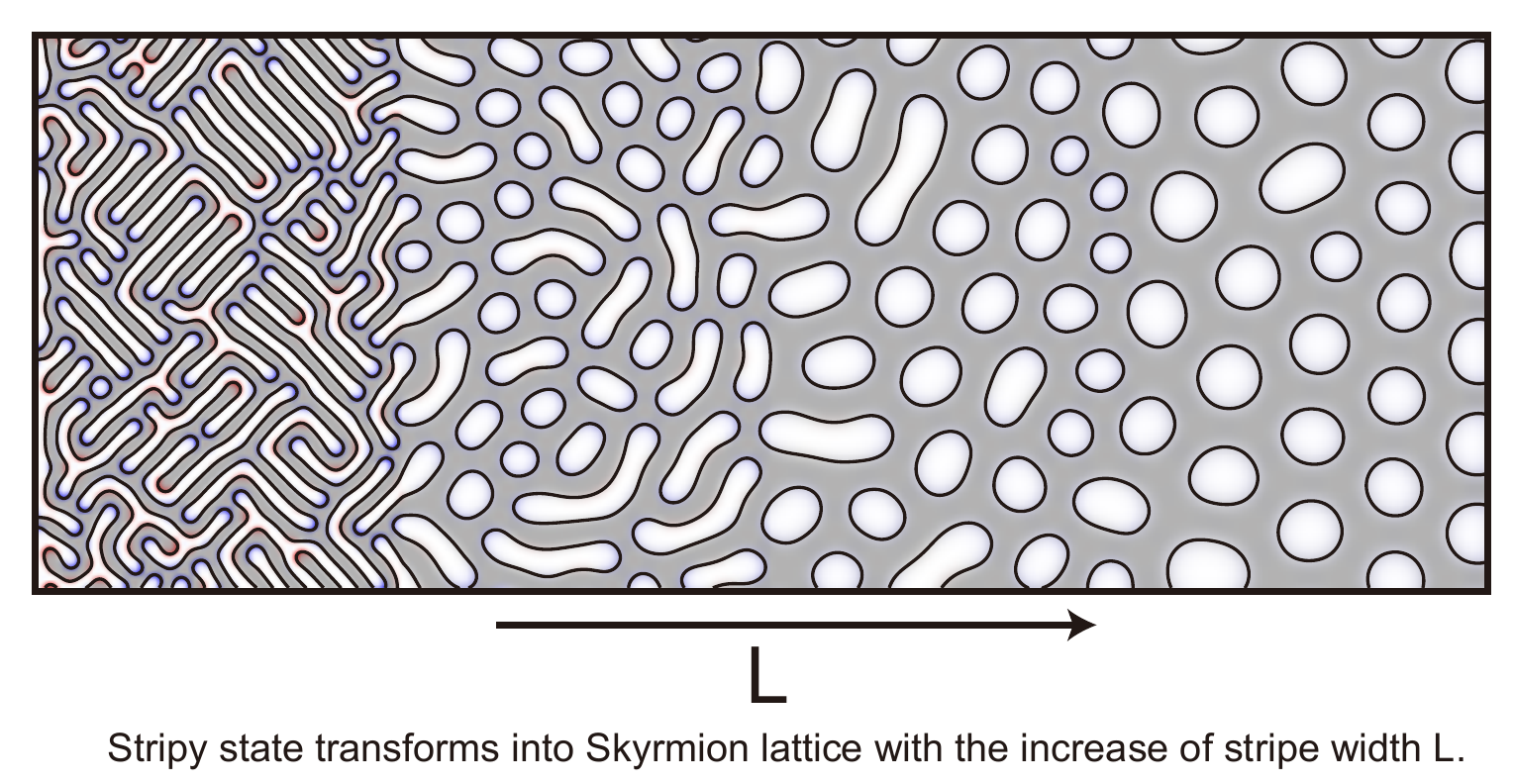}\\

\end{tocentry}

\begin{abstract}
Stripy states, consisting of a collection of stripy spin textures, 
are the precursors of skyrmion crystals (SkXs). Common belief is  
that stripy states and SkXs are topologically unconnected, and 
transitions between SkXs and stripy states are phase transitions. 
Here, we show that both stripy states and SkXs are skyrmion 
condensates and they are topologically equivalent. 
By gradually tuning the stripe whose width goes from smaller 
than to larger than skyrmion-skyrmion separation, the structure 
of a skyrmion condensate transforms smoothly and continuously 
from various stripy phases, including helical states and mazes, 
to crystals, showing that stripy states are topologically 
connected to SkXs. \\ \par
\end{abstract}

Magnetic skyrmions have attracted much attention in recent years for their 
academic interest and potential applications \cite
{Bogdanov2001,Rossler2006,Muhlbauer2009,Yu2010,Yu2011,Nagaosa,Fert,Fert1}. 
Various aspects of magnetic skyrmions have been extensively and intensively 
studied \cite{news,Woo2016,Iwasaki2013,Thiaville2013,Onose,Park,Xiansi,Lenov2016,
Braun,size2016,PdFeIr,koshibas2018,Judge2018,Kim2018,Woo2018,Lin2013,Hoshino2018,
gongxin,gongxin2,MnSi, Zhou2014,Li,Yuan2016,Jiang,Heinze,Xu,MnSi_anis,size2015,
JMMM2,Romming,Tian,Yuan2018,BandK,Jiang2017,Kai2017,
Reich2015,Reich2016,Sampaio2013}. 
Vast experiments \cite{roadmap} showed that various stripy phases, including 
the helical states and mazes, accompany skyrmion crystals (SkXs). 
General belief is that stripy states and SkXs are topologically unconnected 
\cite{roadmap,Karube}, and transformations between them are phase transitions. 
Testing this belief motivates current study. 

For a perpendicularly magnetized chiral film described by magnetization 
$\bf{M}$, the Heisenberg exchange stiffness constant $A$, anisotropic 
coefficient $K$, and the Dzyaloshinskii-Moriya interaction (DMI) coefficient 
$D$, $\kappa\equiv (\pi^2D^2)/(16AK)=1$, an important quantity, separates 
isolated skyrmions ($\kappa<1$) from condensed stripe skyrmions ($\kappa>1$) 
in the absence of a magnetic field \cite{paper1,paper2,paper3,paper4}. 
In this paper, we study how the morphology of a collection of stripe 
skyrmions changes as the stripe widths gradually increase. 
Our results show the topological equivalence of stripy states and SkXs 
and raise the question about the notion of the first-order phase 
transition in SkX formations.  \\ \par

\begin{figure}
	\centering
	\includegraphics[width=8.6cm]{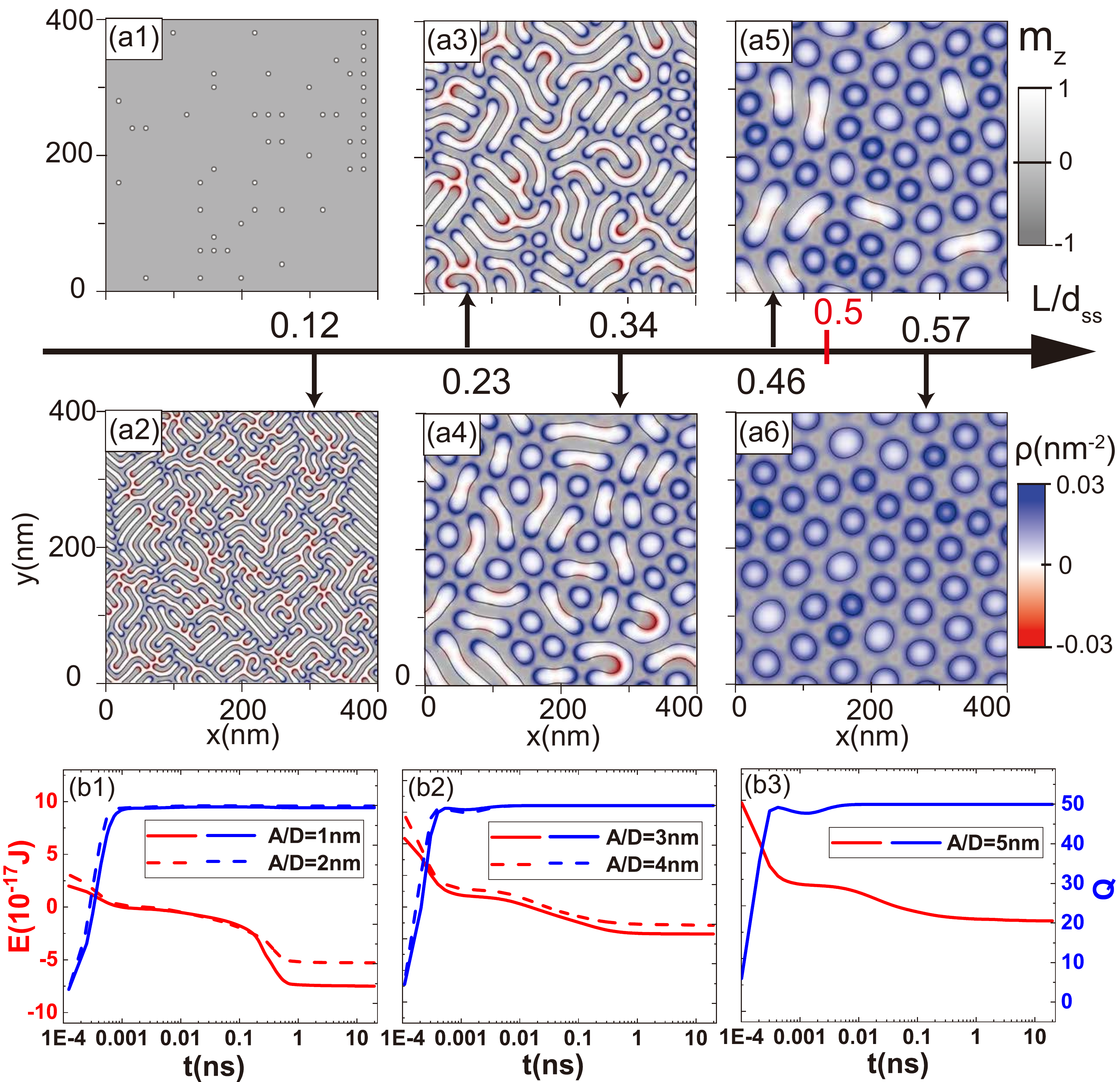}\\
	\caption{(a1) Initial configuration of randomly distributed 50 
	circular nucleation domains of $m_z=1$. The diameter of each 
	domain is $6\, $nm. (a2)-(a6) The stable spin structures with 
	different $A/D$, starting from initial configuration (a1).  
	$A/D=1\, $nm (a2), $2\, $nm (a3), $3\, $nm (a4), $4\, $nm (a5), 
	$5\, $nm (a6), corresponding to Films 1, 3, 5, 7, and 9 in Tab. 
	\ref{table1}. $\kappa=2$ for all films. $L/d_{ss}$ shows the 
	phase-diagram. (b1)-(b3) The time dependences 
	of skyrmion number $Q$ (the right y-axis and the blue curves) and 
	total magnetic energy $E$ (the left y-axis and the red curves). 
	$A/D=1\, $nm, $2\, $nm (b1); $3\, $nm, $4\, $nm (b2); $5\, $nm (b3). 
	The time is in the logarithmic scale. All $Q$ reaches the expected 
	50 within 1 ps, and $E$ approaches its minimum in nanoseconds.}
	\label{fig1}
\end{figure}

We first investigate how morphology of 50 skyrmions, in films of 
$400\,$nm$\, \times 400\, $nm$\, \times 0.5\, $nm, changes with 
stripe width. Previous study \cite{paper1} showed that one 
nucleation domain in a chiral magnetic film of $\kappa=
(\pi D)^2/(16AK)>1$ develops into a stripe skyrmion. 
Thus, 50 nucleation domains of $m_z=1$ embedded in the $m_z=-1$ 
background of chiral magnetic films can be used to generate 50 skyrmions. 
The 50 domains of 6$\, $nm in diameter each are initially distributed in 
films randomly as shown in Fig. \ref{fig1}(a1). Figure \ref{fig1}(a2) 
shows a stable structure of 50 skyrmions for Film 1 with $A/D=1\, $nm (a2). 
All stripes have the same width satisfying $L=a(\kappa)A/D$ with 
$a(\kappa=2)$=6.5 \cite{paper1}. $2L=13\, $nm for Film 1 is much smaller 
than average skyrmion-skyrmion separation $d_{ss}=\sqrt{400\times 400/50}
\simeq 57\,$nm. Since each nucleation domain has a large enough 
space to grow independently, the final stable structure is a 
dense maze formed by 50 ramified/non-ramified stripe skyrmions. 
Figures \ref{fig1}(a3) is a stable structure for Film 3 of $A/D=2\, $nm 
and the dense maze becomes an irregularly arranged 42 stripe skyrmions 
and 8 circular-like skyrmions. Figures \ref{fig1}(a4)-(a6) are stable 
structures for Film 5 with $A/D=3\, $nm (a4); Film 7 with $A/D=4\, $nm 
(a5); and Film 9 with $A/D=5\, $nm (a6). Figures \ref{fig1}(a4)-(a5) are 
the mixtures of 20 circular skyrmions and 30 stripe skyrmions when 
$L=13.02\, $nm (a4), 40 circular skyrmions and 10 stripe skyrmions when 
$L=26.04\, $nm (a5), respectively. SkXs of 50 circular skyrmions form 
when $2L$ is larger than 57$\, $nm as shown in Fig. \ref{fig1}(a6). 
These results demonstrate the importance of $L/d_{ss}$. 
$L/d_{ss}=0.5$ roughly separates SkXs from mixture of stripe skymrions 
and circular skyrmions, as shown in the $L/d_{ss}$ phase diagram. 

Interestingly, both positive and negative skyrmion charges, whose 
density is defined as $\rho=\frac{1}{4\pi}\mathbf{m}\cdot(\partial_x
\mathbf{m} \times \partial_y \mathbf{m})$, appear in a stripe 
skyrmion while circular skyrmions carry only positive charges. 
The colours in Figs. \ref{fig1}(a2)-(a6) encode skyrmion charge 
(positive in blue and negative in red) distribution.
Indeed, each small domain of initial zero skyrmion number becomes 
a skyrmion of $Q=\int\rho\rm dx\rm dy=1$ within 1 picoseconds. 
Each pattern in Figs. \ref{fig1}(a2)-(a6) consists of 50 skyrmions, 
evident from the time-dependence of $Q(t)$ in Figs. \ref{fig1}(b1)-(b3). 
$Q(t)$ reaches $50$ within picoseconds. The total energy $E$ 
monotonically decreases with time and approaches its minimal values 
in nanoseconds for all films as shown in Figs. \ref{fig1}(b1)-(b3) 
(the left y-axis and the red curves). 

\begin{figure}
	\centering
	\includegraphics[width=8.6cm]{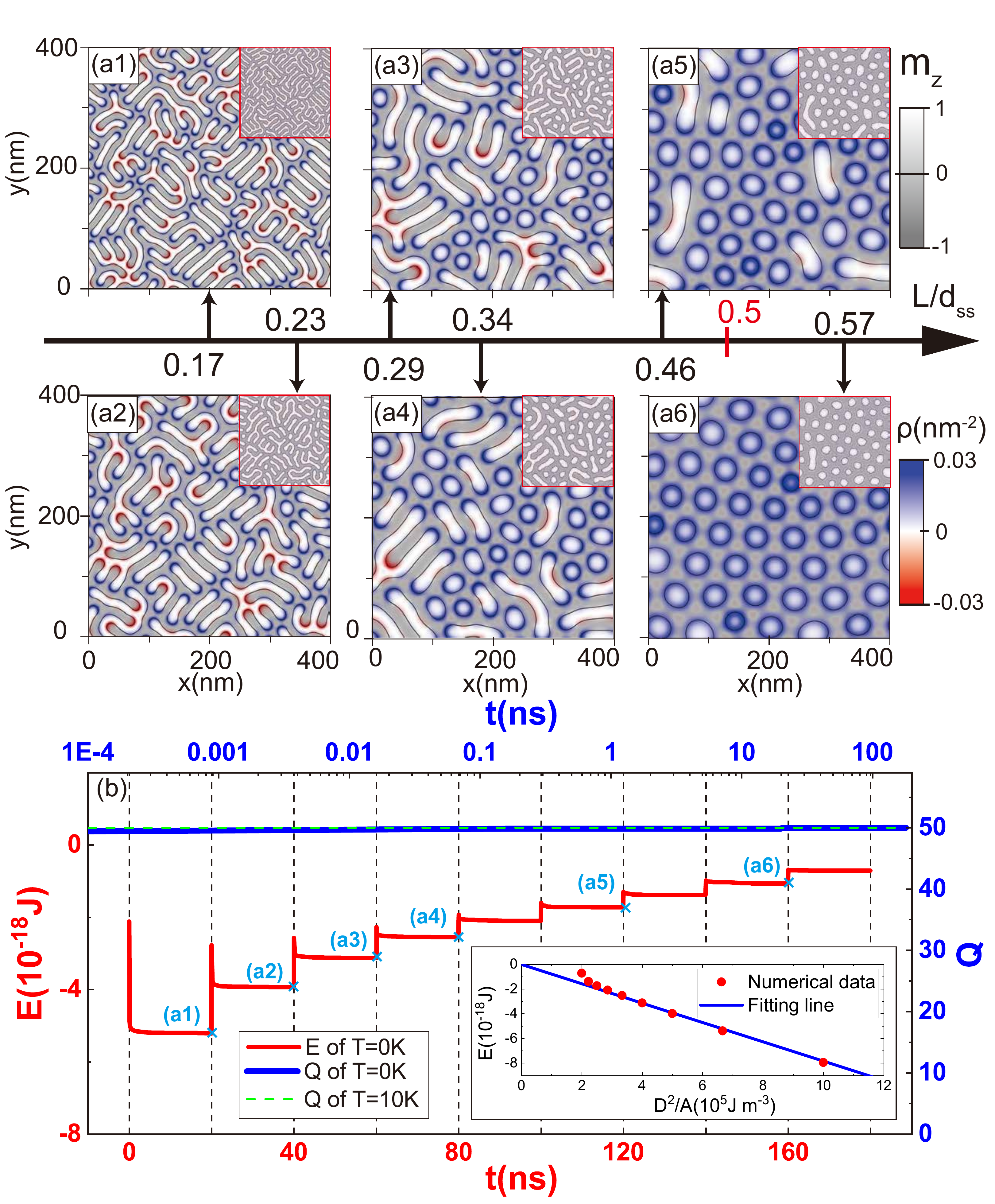}\\
	\caption{(a1) Stable structure of Film 2 developed from the stable structure 
		of Film 1 of Fig. \ref{fig1}(a2). (a2) Stable structure of Film 3 of 
		$A/D=2\, $nm developed from (a1). (a3) Stable structure of Film 4 of 
		$A/D=2.5\, $nm from (a2). (a4) Stable structure of Film 5 of $A/D=3\, $nm 
		from (a3). (a5) Stable structure of Film 7 of $A/D=4\, $nm. (a6) Stable 
		structure of Film 9 of $A/D=5\, $nm. $\kappa=2$ is for all films.  
		Insets: Stable structures at temperature of 10$\,K$ for Film 3 (a2); 
		Film 4 (a3); Film 5 (a4); Film 7 (a5); Film 9 (a6). $L/d_{ss}$ shows 
		the phase-diagram. (b) The blue line ($0\, $K) and the green dash line 
		($10\,$K) are the time evolution of skyrmion number $Q$ (the right y-axis). 
		The time is in the logarithmic scale (up x-axis). 
		The red curve is the total energy $E$ (the left y-axis) as a function of 
		time in the normal scale (down x-axis). $E$ for structures in (a1)-(a6) are 
		marked by cross. $Q$ maintains at 50 for both zero temperature and $10\,K$. 
		$E$ approaches a constant in nanoseconds after tuning the parameters. 
		The inset is total magnetic energy of the stable structure as a function of $D^2/A$.
		The symbols are the simulation results and the blue curve is the fit to 
		$E=c(D^2/A)$ with $c=(-7.8\times10^3\, nm^3)$. }
	\label{fig2}
\end{figure}

In the second set of simulations, starting from the stable maze structure 
of Fig. \ref{fig1}(a2) with 50 ramified stripe skyrmions for Film 1 of 
$A/D=1\, $nm as the initial configuration, we consecutively increase $A$ 
by $0.5\, \rm pJ / m$ and decrease $K_u$ by proper values to keep 
$\kappa=2$ every $20\, $ns to simulate Film 2 to Film 10 listed in Tab. 
\ref{table1}. Figures \ref{fig2}(a1)-(a6) are the stable structures 
for Film 2, 3, 4, 5, 7, and 9, respectively. 
Each of them comes from the stable structure of its predecessor. 
The time-dependences of $E$ and $Q$ are plotted in Fig. \ref{fig2}(b), 
showing that systems reach their equilibrium states (constant $E$ 
and $Q$) within a few nanoseconds with new set of parameters.  
$Q=50$ holds all the time due to the topological nature of the quantity. 


For a given spin structure, the total magnetic energy is, according to Eq. \eqref{energy} 
(see Methods), 
\begin{equation}
	\begin{aligned}
		E &= d\int \int\lbrace A|\nabla\mathbf{m}|^2+D[m_z\nabla\cdot\mathbf{m}\\
		&\quad-(\mathbf{m}\cdot\nabla)m_z]
		+K(1-m_z^2)\rbrace \mathrm{d}x\mathrm{d}y\\
		&= \frac{\pi^2D^2d}{16A}\int\int\lbrace |\frac{4A}{\pi D}\nabla\mathbf{m}|^2+
		\frac{4}{\pi}[m_z\frac{4A}{\pi D}\nabla\cdot\mathbf{m}\\
		&\quad-(\mathbf{m}\cdot\frac{4A}{\pi D}\nabla)m_z]	
		+\frac{1}{\kappa}(1-m_z^2) \rbrace  \mathrm{d}x\mathrm{d}y\\
		&= \kappa K dL^2 \int\int\lbrace |\nabla\mathbf{m}|^2+
		\frac{4}{\pi}[m_z\nabla\cdot\mathbf{m}\\
		&\quad	-(\mathbf{m}\cdot\nabla)m_z]	
		+\frac{1}{\kappa}(1-m_z^2) \rbrace  \mathrm{d}x\mathrm{d}y.
	\end{aligned}	 
	\label{energy1}	
\end{equation} 
Metastable spin structures depend only explicitly on $\kappa=(\pi D)^2
/(16AK)$, and inexplicitly on the length scale of $L_1=4A/(\pi D)$.  
Since the integral area is shrunk by $L_1^2$, the total magnetic energy is 
proportional to $\kappa K \propto D^2/A$ in the case of constant $\kappa$. 
Symbols in the inset of Fig. \ref{fig2}(b) are the total magnetic energies of 
metastable states for $A/D=1\, \rm nm$ up to $5\, \rm nm$. The solid curve 
is the fit to $E(A/D)=cD^2/A$, showing excellent agreement with simulations. 
The energy deviates slightly from $cD^2/A$ for larger $A/D$ such that $L$ 
($L_1$) is comparable to skyrmion-skyrmion separation and skyrmion 
interactions becomes important.

The spikes in $E$-curve come from sudden energy increase when parameters are 
modified. The energy of the new stable structure is higher (less negative) 
as $A$ increases. Compare Fig. \ref{fig2}(a2) with Fig. \ref{fig1}(a3) for 
Film 3, Fig. \ref{fig2}(a4) with Fig. \ref{fig1}(a4) for Film 5, and Fig. 
\ref{fig2}(a5) with Fig. \ref{fig1}(a5) for Film 7, each pair has similar, 
but not identical, stable structures, revealing the fact of a large number 
of metastable structures for a collection of stripe skyrmions. Which 
metastable structure that a mixture of circular and stripe skyrmions ends 
up depends on the history of the system, see more discussions on this 
issue in Supporting Information.

Mumax3 simulations \cite{mumax3} at $T=10\,K$, much lower than the Curie 
temperatures $T_c$ ($84\, $K$\sim 290\,$K for Film 2 to Film 9), are used 
to demonstrate the thermal stability,  see Supporting Information for the 
detail of determining $T_c$. The insets of Fig. \ref{fig2}(a1)-(a6) are the 
metastable structures of Film 2 at 20ns (a1), Film 3 at 40ns (a2), Film 4 
at 60ns (a3), Film 5 at 80$\,$ns (a4), Film 7 at 120$\,$ns (a4), and Film 
9 at 160$\,$ns (a6). $Q=50$ hold throughout 180$\,$ns and morphology of 
the final states are similar as those at zero temperature for all films 
as shown in Fig. \ref{fig2}(b). However, one distinct difference is that 
skyrmion surfaces are more irregular and rough at finite temperatures, 
agreeing with experiments \cite{yzwu}, which may be understood from the 
negative skyrmion formation energy. Negative formation energy means 
negative surface tension such that such a surface can hardly resist 
external perturbations and deformations. 

\begin{figure}
	\centering
	\includegraphics[width=8.6cm]{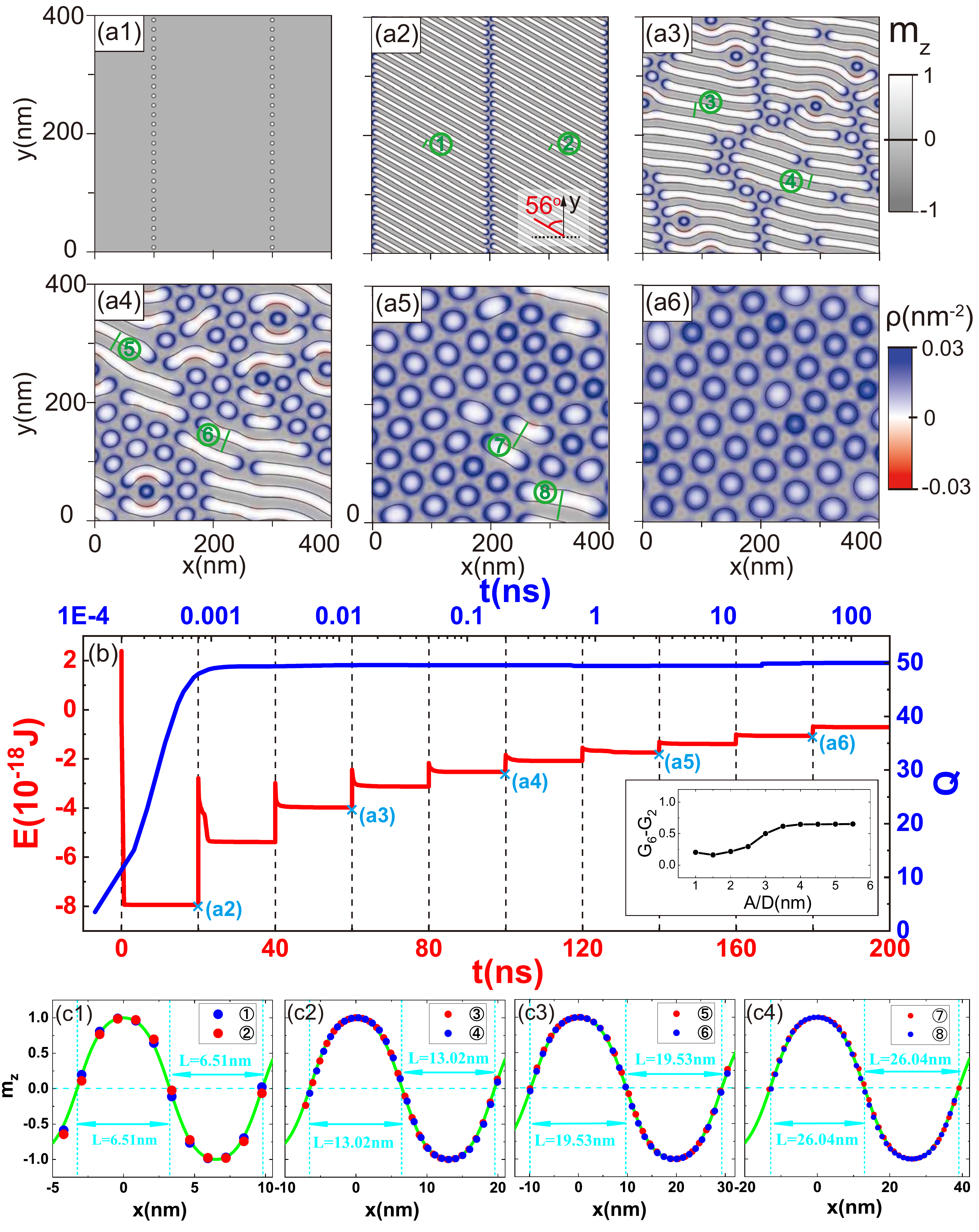}\\
	\caption{
		(a1) Initial configuration of 50 circular nucleation domains of $m_z=1$ 
		arranged orderly in two columns. The diameter of each domain is $6\, $nm.
		(a2)-(a6) Stable structures for $A/D=1\, $nm (a2); $2\, $nm (a3); $3\, $nm 
		(a4); $4\, $nm (a5); and $5\, $nm (a6). All films have $\kappa=2$. 
		(b) The time evolution of skyrmion number $Q$ (the right y-axis and 
		the blue curve). Time is in the logarithmic scale (the top x-axis). 
		The red curve is the total energy $E$ (the left y-axis) as a 
		function of time in the linear scale (the bottom x-axis). 
		$Q$ reaches 50 within 1 ps and keeps this value afterword even 
		when the parameters are tuned to different values. 
		$E$ for structures in (a2)-(a6) are marked by cross. 
		The inset is the $A/D$ dependence of $G_6-G_2$.
		(c1)-(c4) The red and the blue points are the z-component of magnetization 
		along the green lines labeled by circled $n$ in (a2)-(a5). The green lines 
		are the fit to our approximate profile. The cyan arrows indicate the 
		stripe widths of $m_z < 0$ or $m_z > 0$. }
	\label{fig3}
\end{figure}

In contrast to a large number of metastable structures in stripe skyrmion 
condensate, variations of structures in SkXs are limited to a global 
translational transformation of all skyrmions. This explains why SkXs in Figs. 
\ref{fig2}(a6) and \ref{fig1}(a6) are almost the same although their kinetic 
and dynamical process are not the same. To further verify the independence of 
SkXs on their initial configurations, we repeat the simulations for Film 1, 
starting with the same 50 nucleation domains as those in Fig. \ref{fig1} 
but arranged in two well-ordered columns as shown in Fig. \ref{fig3}(a1). 
Figure \ref{fig3}(a2) is the stable helical state after $20\,$ns evolution  
with 50 stripe skyrmions ($Q(t)=50$ as shown in Fig. \ref{fig3}(b)).  
Stripes tilt about $56^\circ$ from the $y$-direction. The stripe tilt 
comes from film tiling by 25 stripes of width $L\simeq 6.5\, $nm. 
The tilted angle $\theta$ satisfies  $2L/(\sin \theta)=400/25$. 
For $L=6.5\,$nm, $\theta=\cos^{-1}(13/16)\simeq 54^\circ$, very close 
to the observed value of $56^\circ$. 

After consecutively changing model parameters listed in Tab. \ref{table1} 
for Film 3 to 10 every 20$\, $ns, Figs. \ref{fig3}(a3)-(a6) show stable 
structures for Film 3 with $A/D=2\, $nm at $60\, $ns (a3), Film 5 with 
$A/D=3\, $nm at $100\, $ns (a4), Film 7 with $A/D=4\, $nm at $140\, $ns (a5), 
and Film 9 with $A/D=5\, $nm at $180\, $ns (a6). Figure \ref{fig3}(b) shows 
time-dependences of $E$ and $Q$. The spikes in $E$-curve are from the sudden 
change of model parameters. $Q$ maintains the value of 50 after the first 
a few picoseconds while $E$ reaches its minimal value in nanoseconds every 
time after parameter changes. The number of stripe (circular) skyrmions 
decreases (increases) as $L$ increases. All skyrmions in Fig. \ref{fig3}(a6) 
are circular when $L$ is larger than the average skyrmion-skyrmion separation.

To demonstrate that structure transformation from stripy states to SkXs is 
smooth, we introduce a local orientation order parameter, widely used in 
studying triangular lattices \cite{KTNHY1,KTNHY2,KTNHY3}, for $i$'th skyrmion  
\begin{equation}
	\Psi_n(i)=\sum_{NN}e^{in\theta_{ij}}.
	\label{order1}
\end{equation} 
$\theta_{ij}$ is the angle of the direction from the center of skyrmion 
$i$ to the center of skyrmion $j$ with respect to the x-axis, and sum is 
over all nearest neighbouring skyrmions $j$. For a lattice with orientation 
order of a triangular lattice, $\Psi_6(i)$ is close to 1 while it is a small 
complex number for a lattice without orientation order. For an imperfect 
lattice structure, averaged $\Psi_n$ (over all skyrmions), called $G_n$, 
\begin{equation}
	G_n=\frac{1}{N}\sum_{i}\Psi_n(r_i)\Psi_n^*(r_i),
	\label{order2}
\end{equation} 
can distinguish one skyrmion structure from another \cite{KTNHY1,KTNHY2,KTNHY3}. 
$G_6$ for a SkX in a perfect triangular lattice is 1.   
Since $G_6$ is also close to 1 for ordered stripe states such as those in Figs. 
\ref{fig3}(a2)-(a3), we use $G_6-G_2$, which is close to 0 for a helical state 
and 1 for a SkX, as an order parameter to distinguish the two structures.
$G_6-G_2$ undergoes a sudden transition for a first-order phase transition 
and a smooth variation for a non-phase transition. The inset of Fig. 
\ref{fig3}(b) shows how $G_6-G_2$ varies with $A/D$ from 0 for an ordered 
stripe skyrmion state, via maze and mixture of stripes and circular skyrmions, 
to 1 for SkXs. The smooth variation of $G_6-G_2$ further supports the 
assertion of no first-order phase transition from helical states to SkXs. 
It should be pointed out that our SkXs, similar to those in the literature, 
are not in perfect triangular lattice and $G_6-G_2$ is smaller than 1. 

Previous studies \cite{paper1,paper2,paper3,paper4} showed that stripe 
skyrmion structure follows $m_z(x)=\frac{\sinh^2(L/2w)-\sinh^2(x/w)}
{\sinh^2(L/2w)+\sinh^2(x/w)}$ for $-L/2<x< L/2$ and $m_z(x)=-\frac{
\sinh^2(L/2w)-\sinh^2(x/w-L/w)}{\sinh^2(L/2w)+\sinh^2(x/w-L/w)}$ for 
$L/2<x<3L/2$, where $L$ and $w$ are the stripe width and wall thickness. 
$x=0$ refers to stripe center where $m_z= 1$.  
Figures \ref{fig3}(c1)-(c4) are the distributions of $m_z$ along lines  
labelled by the green \textcircled{n} in Figs. \ref{fig3}(a2)-(a5). 
The solid curves are the fits to the theoretical spin profile with  
$L=6.51\, $nm and $w=1.73\, $nm for (c1); $L=13.02\, $nm and $w=3.46\, $nm 
for (c2); $L=19.53\,$nm and $w=5.19\, $nm for (c3); $L=26.04\, $nm and 
$w=6.92\, $nm for (c4). $L$ agrees perfectly with $L=6.51A/D$ for $\kappa=2$.
Data from different stripes falling onto the same curve demonstrates that 
stripes, building blocks of stripy pattern, are identical.
Although the initial configurations in Figs. \ref{fig1}, \ref{fig2}, and 
\ref{fig3} are different, and the morphologies of the mixtures of stripe 
and circular skyrmions are sensitive to the initial configurations and 
their dynamical paths, the structures of SkXs do not depend on the dynamical 
paths and depend only on the ratio of the stripe width and skyrmion-skyrmion 
separation. 

\begin{figure}
	\centering
	\includegraphics[width=8.6cm]{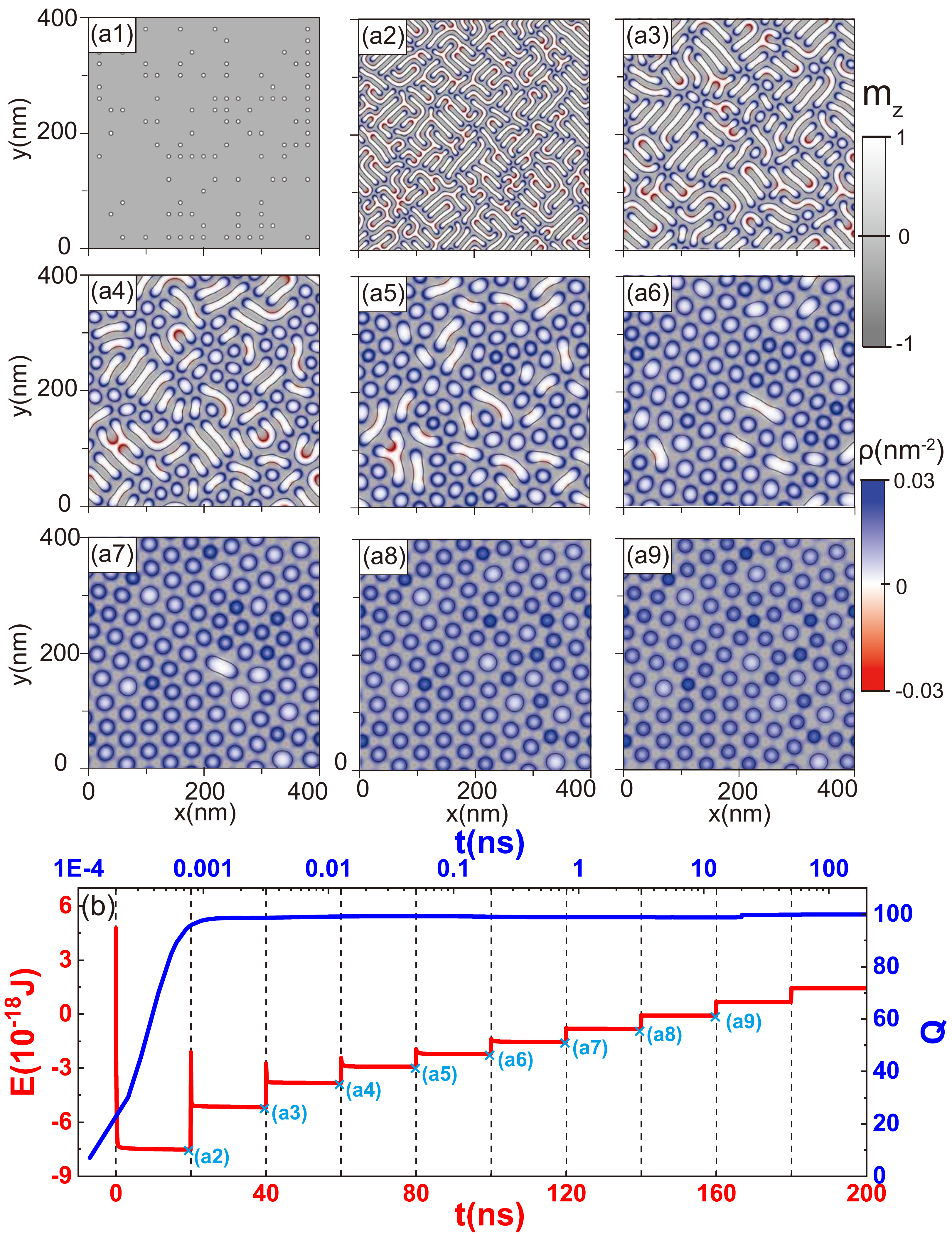}
	\caption{(a1) The initial configuration with 100 randomly distributed nucleation 
		domains of $m_z=1$ in the films. The diameter of each domain is $6\, $nm.
		(a2)-(a9) The stable structures for $A/D=1\, $nm at $t=20\, $ns (a2); $1.5\, $nm 
		at $t=40\, $ns (a3); $2\, $nm at $t=60\, $ns (a4); $2.5\, $nm at $t=80\, $ns (a5); 
		$3\, $nm at $t=100\, $ns (a6); $3.5\, $nm at $t=120\, $ns (a7); $4\, $nm at 
		$t=140\, $ns (a8); $4.5\, $nm at $t=160\, $ns (a9). $\kappa=2$ is for all films.   
		(b) The time-dependences of skyrmion number $Q$ (the right y-axis and the blue curve) 
		and the total energy $E$ (the left y-axis and the red curves). Time $t$ is in the 
		linear scale for $E$ (the lower x-axis) and in in the logarithmic scale (the top 
		x-axis) for $Q$. $E$ for structures in (a2)-(a9) are marked by cross. $Q$ reaches 
		100 within 1 ps and remains this value even when model parameters are tuned. 
		$E$ approaches its minimal value in nanoseconds after model parameters are tuned.}
	\label{fig4}
\end{figure}

To further substantiate the assertion above, we carry out similar simulations 
as those for Fig. \ref{fig2}, but with 100 randomly distributed nucleation 
domains as the initial configuration. The average skyrmion-skyrmion separation 
is reduced to $d_{ss}=\sqrt{400\times 400/100}=40\, $nm. Thus, we should 
expect the transformation from maze to the mixture of stripe and circular 
skyrmions, and to a SkX at smaller $A/D$ than those in Fig. \ref{fig2}.  
As shown in Fig. \ref{fig4}, this is indeed the cases. (a1) is the initial 
configuration of randomly distributed 100 circular nucleation domains. 
(a2) is the stable structure of Film 1 of $A/D=1\, $nm and $\kappa=2$ after 
$20\, $ns evolution. Clearly, it is a maze of 100 skyrmions, evident from 
$Q=100$ within a few picoseconds of evolution, and individual skyrmion is 
an irregular ramified stripe, or a curved or short stripe. 
After consecutive changes of $A$ and $K$ to simulate Films 3 to 10 given 
in Tab. \ref{table1} every $20\, $ns, Figs. \ref{fig4}(a3)-(a9) are their 
stable structures at $t=40 \, $ns; $60 \, $ns; $80\, $ns; $100\, $ns; 
$120\, $ns; $140\, $ns; and $160\, $ns respectively for $A/D=1.5\, $nm (a3); 
$A/D=2\, $nm (a4); $2.5\, $nm (a5); $3\, $nm (a6), 3.5$\, $nm (a7), 4$\, $nm 
(a8), and 4.5$\, $nm (a9) with stripe width $L=9.76\, $nm, $13.02\, $nm, 
$16.27\, $nm, $19.53\, $nm, $22.78\, $nm, $26.04\, $nm, $29.29\, $nm, 
$32.55\, $nm. Indeed, 100 skyrmions form SkXs when $A/D>3\, $nm as shown in 
Figs. \ref{fig4}(a7)-(a9), smaller than $4\,$nm in Fig. \ref{fig2}. 
For $A/D=3\, $nm, almost all skyrmions are circular except 9 very short 
stripe skyrmions (a6). Figure \ref{fig4}(b) is the time dependences of total 
energy $E$ and $Q$ to show the steady states. $Q(t)$ demonstrates clearly the 
topological feature of the quantity and its resilience to environmental changes. 

Our results can either be experimentally tested or be used to explain 
experiments because magnetic material parameters depend on the temperature. 
Experiments \cite{A-D-T1,A-D-T2} showed that $A$ and $D$ depends on $M$ 
very differently although mean field theories predict the same dependence 
of $A$ and $D$ on $M$. Of course, both $L$ and $\kappa$ change with the 
temperature in realities \cite{A-D-T1,A-D-T2}, in contrast to the assumption 
of constant $\kappa$ here. Since skyrmion structures are determined by the 
ratio of $L$ and skyrmion-skyrmion separation, not on $\kappa$ as long 
as it supports condensed skyrmion phase ($\kappa>1$) such that whether 
$\kappa$ is a constant or not is not essential. Beside the temperature, 
other external knobs such as electrical fields, magnetic fields, as well 
as strains can also change $L$ \cite{Schott2021,Srivastava2018,Deger2020,
Gareeva2020,Udalov2020}. 

Physics reported here occurs in the bulk, and should not depend on the 
boundary conditions. This is also what observed in experiments \cite{yzwu}.  
However, it is known that helical spin structures ended at a sample edge 
can be very complicated. Skyrmion numbers of such spin textures are 
usually not integer. In the Supporting Information, we discuss possible 
features of textures at sample edges. 

Looking forward, our assertion of no first-order phase transition may not 
be applicable to SkX formation from the closely packed isolated skyrmions. 
The similarities and differences between the two types of SkXs 
should be an interesting issue that deserves a careful study. 
SkXs from a collection of isolated skyrmions due to the 
skyrmion-skyrmion interaction has many similarities with the 
formation of atomic crystals through atom-atom interactions. 
It is well known that the melting or crystallization of an atomic 
crystal to or from a liquid or an amorphous 
state is the first-order phase transition involving latent heat. 
Of course, unlike real crystals in which atoms cannot be destroyed, 
skyrmion number does not need to be conserved in SkX formation.

In conclusion, a collection of skyrmions transforms into a SkX as the width 
of stripe skyrmions becomes larger than the skyrmion-skyrmion separation, 
no matter whether the initial configuration of skyrmions are ordered or 
randomly distributed in a film and whether the width increases gradually 
or suddenly. Since the SkX comes from the skyrmion-skyrmion repulsion, it 
explains why individual skyrmion in a SkX is not a truly circular spin texture. 
Our finding does not support the notion of the first-order phase transition 
for SkX formation, but current results cannot rule out the second-order 
or higher-order phase transitions. 

\section{Methods}

Our model is a ferromagnetic film of thickness $d$ in $xy$ plane with 
the total magnetic energy $E$ consisting of the exchange energy 
$E_\mathrm{ex}$, the DMI energy $E_\mathrm{DM}$, the anisotropy 
energy $E_\mathrm{an}$, and the magnetic dipolar energy $E_\mathrm{d}$
\begin{equation}
	E=E_\mathrm{ex}+E_\mathrm{DM}+E_\mathrm{an}+E_\mathrm{d},
	\label{energy}
\end{equation}
where $E_\mathrm{ex}=Ad\iint |\nabla \mathbf{m}|^2\mathrm{d}S$, $E_
\mathrm{DM}=Dd\iint [m_z\nabla\cdot\mathbf{m}-(\mathbf{m}\cdot\nabla)
m_z]\mathrm{d}S$, $E_\mathrm{an}=K_ud\iint (1-m_z^2)\mathrm{d}S$, and 
$E_\mathrm{d}=\mu_0 M_{\rm s}d\iint\mathbf{H}_{\rm d}\cdot\mathbf{m}
\mathrm{d}S$. $M_{\rm s}$ and $\mathbf{m}$ are the saturation 
magnetization and the unit direction of $\mathbf{M}$, respectively. 
$K_u$ is the perpendicular magneto-crystalline anisotropy and 
$\mu_0$ is the vacuum permeability. Magnetization dynamics is 
governed by the Landau-Lifshitz-Gilbert (LLG) equation,
\begin{equation}
	\frac{\partial\mathbf{m}}{\partial t}=-\gamma\mathbf{m} \times\mathbf{H}
	_{\rm eff}+\alpha\mathbf{m}\times\frac{\partial\mathbf{m}}{\partial t},
	\label{llg}
\end{equation}
where $\gamma$ and $\alpha$ are gyromagnetic ratio and the Gilbert 
damping constant, respectively. $\mathbf{H}_{\rm eff}=\frac{2A}{\mu_0 M_s}\nabla^2
\mathbf{m}+\frac{2K_u}{\mu_0 M_s}m_z\hat z+\mathbf{H}_{\rm d}+ \mathbf{H}_{\rm DM}+
\mathbf{h}$ is the effective field including the exchange field, crystalline 
anisotropy field, demagnetization field, DMI field, and a temperature-induced random 
magnetic field of magnitude $h=\sqrt{2\alpha k_{B} T/(\mu_0 M_s \gamma \Delta V 
\Delta t)}$ where $\Delta V$, $\Delta t$, and $T$ are the cell volume, time step, 
and the temperature, respectively. Demagnetization field can be included in the 
effective anisotropy $K=K_u-\mu_0M_s^2/2$ when the film thickness $d$ is much 
smaller than the exchange length \cite{Xiansi}. The thickness of our samples are 
$0.5\, $nm which is indeed much smaller than the exchange length of 
$\sqrt{2A/(\mu_0M_s^2)}>60\, $nm, where $\mu_0=4\pi\times 10^{-7}\,{\rm N/A^{2}}$
is vacuum permeability. 

In the absence of energy sources such as an electric current and the 
heat bath, the LLG equation describes a dissipative system whose energy 
can only decrease \cite{xrw1}. Thus, the steady state solutions 
of LLG equation with proper initial magnetization distributions 
are the stable/metastable spin textures of Eqs. \eqref{energy}. 
This is an efficient way to find the structure of a skyrmion condensate.
Mumax3 package \cite{mumax3} is used to search various metastable spin 
textures in films of $400\, \rm nm\times\, 400 \, nm\times \, 0.5\, nm$ 
with periodic boundary conditions in both $x$ and $y$-directions. 
The mesh size is $1\, \rm nm \times\, 1\,nm\times \, 0.5\, nm$ so that no 
change in simulation results is detected when a smaller mesh size is used. 
In this study, we consider 10 different films with the same 
$M_s=0.15{\,} \rm MA/m$ and $D=1\, \rm mJ/m^{2}$. 
$A$ and $K_u$ can vary and are used 
to tune the stripe width while $\kappa=2$ are the same for all 10 films. 
The values of $A$ and $K_u$, as well as the effective $K$ are listed in Tab. 
\ref{table1}. $A/D$ of the 10 films varies from 1$\, $nm to 5.5$\, $nm. 
Also a large $\alpha=0.5$ is used to speed up our simulations. 

\begin{table}[htbp]
	\centering
	\setlength{\tabcolsep}{2mm}{
		\caption{The value of material parameters, $A$, $K_u$, and $K$ for 
			10 different films.		}
		\label{table1}	
		\begin{tabular}{llll}
			\hline\hline\noalign{\smallskip}
			Films &$A (\, \rm pJ/m\,)$
			&$K_u (\, \rm MJ/m^{3}\, )$
			&$K (\, \rm MJ/m^{3}\, )$
			\\ \noalign{\smallskip}\hline\noalign{\smallskip}
			1 & 1.0  & 0.322 & 0.308        \\ 
			2 & 1.5  & 0.220 & 0.205        \\ 
			3 & 2.0  & 0.168 & 0.154       \\ 
			4 & 2.5  & 0.137 & 0.122       \\
			5 & 3.0  & 0.116 & 0.102        \\
			6 & 3.5  & 0.102 & 0.088       \\
			7 & 4.0  & 0.091 & 0.077        \\
			8 & 4.5  & 0.082 & 0.068        \\
			9 & 5.0  & 0.075 & 0.061        \\
			10 & 5.5 & 0.070 & 0.056        \\
			\noalign{\smallskip}\hline\hline
	\end{tabular} } 	
\end{table} 

\section{ASSOCIATED CONTENTS}

\noindent{\bf Supporting Information}\\ \par
The Supporting Information is available free of charge on the ACS 
Publication website at DOI:
	
\begin{itemize}
\item Supporting Information: Discussion of spin structures at sample boundary, 
damping dependence of metastable state, and determination of the Curie 
temperatures. (PDF)

\end{itemize}

\section{AUTHOR INFORMATION}
\par
\noindent{\bf Corresponding Author}\\
*E-mail: phxwan@ust.hk. Phone: +852-23587488. Fax: +852-23581652.\\

\noindent{\textbf{Notes}}\\
The authors declare no competing financial interest.\\
\par
\section{ACKNOWLEDGMENTS}
\par
This work is supported by the National Key Research and Development Program 
of China (grant No. 2020YFA0309600), the NSFC (grant No. 11974296), and 
Hong Kong RGC Grants (No. 16300522, 16301619 and 16302321). 
\\ \par

\end{document}